\documentclass[aps,prl,reprint,groupedaddress,showpacs,amsmath, amssymb, floatfix]{revtex4-1}
\usepackage{epstopdf}
\usepackage{graphicx}
\usepackage{color}
\usepackage{latexsym}
\usepackage{amssymb}
\usepackage{bm}
\usepackage{dcolumn}
\usepackage{soul}

\newcommand{\he}{$^3$He}

\begin{document}
\title{New Limits on Transverse Zero Sound in Fermi Liquid \he}

\author{M. D. Nguyen}
\email{mannguyen2019@u.northwestern.edu,\\ w-halperin@northwestern.edu}
\author{D. Park}
\author{J. W. Scott}
\author{N. Zhelev}
\author{W. P. Halperin}
\affiliation{Department of Physics and Astronomy, Northwestern University \\Evanston, IL 60208, USA}

\date{\today}

\begin{abstract}
Landau predicted that transverse sound propagates in a Fermi liquid with sufficiently strong interactions such as for liquid \he, unlike a classical fluid which cannot support shear oscillations. Previous attempts to observe this unique collective mode yielded inconclusive results masked by single particle excitations. We  micro-fabricated acoustic cavities with a micron scale path length that is suitable for direct detection of this sound mode, and used the interference fringes from our acoustic Fabry-P\'erot cavities to determine both real and imaginary parts of the acoustic impedance.  No interference was observed. Either transverse sound does not exist or its attenuation must be very high, well above 2000\,cm$^{-1}$. We discuss a theoretical framework for such high attenuation.
\end{abstract}

\maketitle

\section{Introduction}

The Fermi liquid model, proposed by Landau, \cite{lan.57a} and expanded upon later by many others, lies at the foundation of theoretical condensed matter physics and our understanding of the Fermi liquid state. Nearly all the central predictions of Landau's model have been verified, including the existence of longitudinal zero sound in liquid \he \,\cite{abe.66}. Landau also predicted the existence of another propagating sound mode, called transverse zero sound (TZS), where the polarization of the wave is perpendicular to the wavevector. Measurements by Roach \textit{et al.} \cite{roa.76} were initially believed to demonstrate the existence of TZS in \he\, but were later shown by Flowers \textit{et al.} \cite{flo.76} to be the result of an incoherent quasiparticle excitation. 

It was originally expected that transverse sound would be more difficult to detect in the superfluid state as the number of unpaired quasiparticles reduces with temperature. However, it was shown by Moores and Sauls that a coupling to the collective modes of the superfluid state contributes to the stress tensor in a manner much stronger than Landau's predicted mode in the degenerate Fermi liquid\cite{moo.93}. This coupling supports a propagating transverse sound mode which exhibits an acoustic Faraday effect.  Both were experimentally demonstrated by Lee \textit{et al.}\,\cite{lee.99}.  That work led to the discovery of a new order parameter collective mode\cite{dav.08b}. At this date, superfluid \he\  remains the only fluid where transverse sound waves have been observed. In the present work we revisit the normal Fermi liquid state to search for the last unverified prediction of Landau, using the results in the superfluid as a quantitative reference.

\begin{figure}
\includegraphics[width=8.2cm]{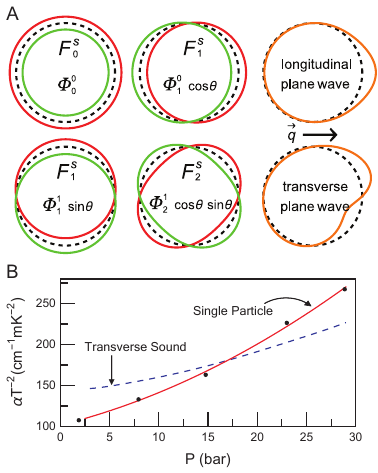}
\caption{ {\bf Landau's prediction and previous experiment.}\\({\bf A}) Oscillations of the Fermi surface for various $l$-channels and sound modes for a rightward moving plane wave solution to Eq. \ref{eq:waveEq} at T = 0. ({\bf B})  Flowers \textit{et al.} \cite{flo.76} showed that the pressure dependence of the attenuation reported by Roach \textit{et al.} \cite{roa.76} is consistent with  single particle contributions; not transverse sound.
}
\label{fig:F1}
\end{figure}

In a Fermi liquid, low-energy quasiparticle excitations near a degenerate Fermi surface respond to external fields, leading to changes in the quasi-particle distribution function, $n_{\bm{k}}$. Landau showed that the response of the Fermi liquid, determined by the interactions of these quasiparticles, can be modeled as distortions of the Fermi surface in momentum space, as seen in Fig.\ref{fig:F1}A. The Fermi surface can be thought of as a vibrating membrane and the interactions can be projected onto a basis of different angular momentum channels for the corresponding distortions of the Fermi surface.

The dynamics of wave propagation in a Fermi liquid are governed by the Landau-Boltzmann kinetic equation \cite{lan.57b}:
\begin{equation}\label{eq:waveEq}
i\left( v_F ~ \bm{\hat{k}}\cdot\bm{q}  - \omega\right)\Phi_{\bm{\hat{k}}} + i ~ v_F ~ \bm{\hat{k}}\cdot\bm{q} ~ \int \frac{d \Omega_{\bm{k'}}}{4 \pi} \mathbb{F} (\bm{\hat{k}}\cdot \bm{\hat{k}'}) \Phi_{\bm{\hat{k}'}} = \delta I_{\bm{\hat{k}}},
\end{equation}
where $v_F$ is the Fermi velocity, $\bm{q}$ and $\omega$ are the wave-vector and the frequency of sound, $\Phi_{\bm{\hat{k}}}$ encodes the variation of the Fermi surface, where  $\mathbb{F} (\bm{\hat{k}}\cdot \bm{\hat{k}'})$ are the Fermi liquid interactions between quasiparticles with momentum $\bm{k}$ and $\bm{k'}$ , and $\delta I_{\bm{k}}$ is the linearized collision integral for binary quasiparticle collisions. The Fermi liquid interactions can be projected in  different angular momentum channels for spin-symmetric and spin anti-symmetric interactions. Only spin-symmetric interactions are relevant for sound propagation. The strength of each $l$-channel is given by the Fermi liquid parameters $F_l^s$; $s$ denotes spin symmetric. 

In the hydrodynamic regime, the quasiparticle collision rate, $1/\tau$, is much higher than the frequency of sound so $\omega \tau \ll 1$.
At low temperatures  well below the Fermi temperature, $T \ll T_F$, the collision rate is low and  insufficient to restore oscillations to equilibrium. The Fermi liquid interaction dominates and the sound mode is collisionless  zero sound, $\omega \tau \gg 1$. Landau predicted that for sufficiently strong values of $F_l^s$, not only longitudinal zero sound but also transverse zero sound, can exist. 

While any positive, non-zero value of $F_1^s$ leads to transverse current fluctuations, a transverse sound wave will only propagate if the speed of sound relative to the Fermi velocity is $c_t/v_F > 1$, imposing a condition on the Fermi liquid parameters, \cite{fom.68,flo.78}
\begin{equation}\label{eq:condition}
\frac{F_1^s}{3} + \frac{F_2^s}{1+F_2^s/5} > 2,
\end{equation}
a condition satisfied by liquid \he\ at  all pressures. Therefore TZS can be expected to propagate. 

\begin{figure}
\includegraphics[width=8.2cm]{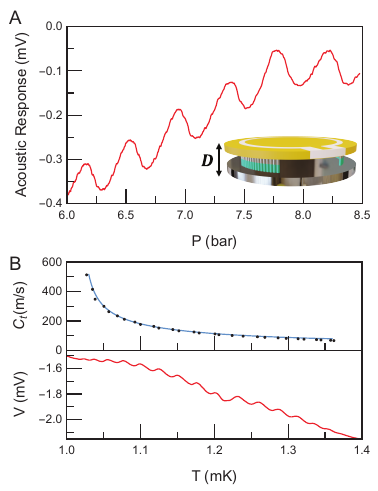}
\caption{{\bf Transverse sound in Superfluid \he.}\\({\bf A}) Interference fringes obtained during a pressure sweep in the superfluid state at 0.95 mK. Transverse sound results from off-resonant coupling to the $J=2-$ Higgs mode in the superfluid \cite{ngu.19,sau.22}. The inset shows the acoustic cavity formed between a piezoelectric AC-cut quartz transducer (top, gold disk) and a micro-fabricated silicon reflecting plate (bottom disk).  ({\bf B}) Interference fringes obtained during a temperature sweep at 10.5 bar and 103.9 MHz in the superfluid state are shown  used to calculate the speed of sound showing good agreement with the theory of Moores and Sauls \cite{moo.93} for a cavity size of $D$\,=\,5.2 $\mu$m.}
\label{fig:F2}
\end{figure}

Roach \textit{et al.} measured the transverse acoustic impedance of  \he\  as a function of temperature from the hydrodynamic regime  ($\sim$ 100 mK) into the zero sound regime ($\sim$ 2 mK) \cite{roa.76}, with features that were thought to be consistent with zero sound; however, this was shown to be otherwise by Flowers {\it et al.} \cite{flo.76}, as displayed in Fig.\ref{fig:F1}B.

In order to distinguish between TZS and an incoherent quasiparticle excitation, an interference experiment must be performed. In the present work we have significantly extended the sensitivity range of acoustic Fabry-P\'erot interferometers that have been used previously to study transverse sound in the superfluid state \cite{lee.99,dav.08a,col.15} (Supplementary Materials).  Importantly, our limiting low temperature and pressure-independent measurements of attenuation in the superfluid, $\sim\,500$ cm$^{-1}$, allow us to calculate the expected attenuation of TZS in the normal Fermi liquid.

We used an acoustic cavity formed between a piezoelectric transducer and a reflecting plate,  shown in Fig.\ref{fig:F2}A, details in Supplementary Materials. Depending upon the speed of sound and path length $D$, the reflected signal will interfere destructively or constructively  at the transducer surface, creating an oscillating signal corresponding to interference fringes in the acoustic impedance as the velocity of sound is varied by sweeping pressure or temperature. In this figure for the superfluid, each oscillation corresponds to a change of one wavelength in a distance of $2D$.  The amplitude of the interference pattern is proportional to the attenuation coefficient, $\delta \mathcal{A} = e^{-\alpha 2D}$, easily observed in the superfluid \he\, B-phase with cavities of $\sim$ 30$\,\mu$m.  Using micro-fabrication techniques we have reduced $D$  to  5.2$\,\mu$m  to  increase signal sensitivity by $\sim$\,400\, in order to accommodate  expected high attenuation in the normal Fermi liquid, $\mathcal{O} (1000 $ cm$^{-1})$\cite{lea.73}.

\begin{figure}
\includegraphics[width=8.2cm]{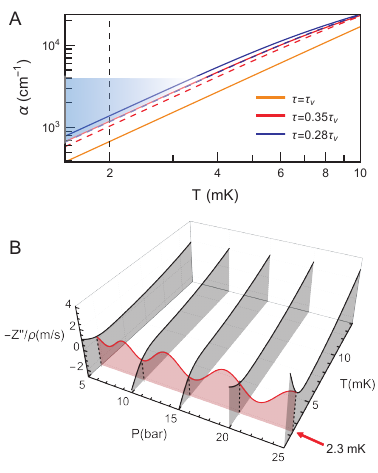}
\caption{{\bf Simulation}\\
({\bf A}) Calculated attenuation coefficient, $\alpha$, for TZS as a function of temperature, for different relaxation times,  $\alpha \sim T^2$. Measurements were at $\sim$\,2 mK where temperatures are in the zero sound regime in the Fermi liquid below 5 mK, blue shaded region. Solid and dashed lines are  for $F_2^s\,= 0$ and 1, respectively.({\bf B}) Imaginary part of the transverse impedance $Z''$ (Eq. 3) under a pressure sweep at 2.3 mK (red line) and temperature sweep at various fixed pressures (black lines). Parameters  in the simulation were tabulated in Ref \cite{hal.90b}, except the viscous relaxation time $\tau_v$ taken from Wheatley\cite{whe.75}. A diffusive boundary condition was used. The simulated attenuation at 2.3 mK was  $\sim\,1000\,\text{cm}^{-1}$.}\label{fig:F3}
\end{figure}

The number of interference fringes is  controlled by changes in the speed of sound. 
Its pressure dependence occurs separately in $v_F$ and $c_t$. The Fermi velocity, $v_F$,  decreases from 59 m/s at 0 bar to 32 m/s at 34.39 bar and $c_t$ varies less with pressure than $v_F$. Constraints on high frequency propagation are discussed in Supplementary Materials.  Nonetheless, interference fringes from TZS should be well within the sensitivity range of the transducers used for a cavity of 5 $\mu$m operated at $\sim$\,100 MHz.

\section{Results} 
We first consider experiments in the superfluid  to characterize the quality of the cavities and calibrate sensitivity in measurement of attenuation.  The amplitude of the interference fringes is proportional $\delta \mathcal{A}$. If the amplitude of interference can be determined for a known value of $\alpha$, then it can be used to estimate the cavities detection sensitivity to highly attenuated signals. From past experiments, the smallest $\delta \mathcal{A}$ detected was $\sim$ 0.03 \cite{dav.08d}. 

However, this estimate does not include surface Andreev bound states that contribute to attenuation  \cite{dav.08d} . Otherwise, $\alpha$ is expected to decrease to zero at T = 0, and there could be attenuation from spontaneous emission due to vacuum fluctuations. We found that $\alpha$ saturates around 400 to 500 cm$^{-1}$, independent of pressure\cite{dav.08d}, and there is no observed decrease in $\alpha$ below 0.5 $T_c$ which provides a good background reference  for the present work. By cooling to $\sim$\, 0.5 $T_c$ and performing a pressure sweep, one can obtain the expected amplitude of the interference fringe for transverse sound in the superfluid propagating with 500 cm$^{-1}$ of attenuation, where $T_c$ is the superfluid transition temperature. This provides a quantitative reference  for the background to the attenuation measurements in the Fermi liquid. 

We used both temperature and pressure sweeps to calibrate our cavities. A demagnetization cooling temperature sweep at 103.9 MHz and 10.5 bar is shown in Fig. \ref{fig:F2}\,B. Changing temperature affects the superfluid gap $\Delta$, the Tsuneto function \cite{tsu.60}, and the $J=2-$ Higgs order parameter collective mode frequency in a known way, Nguyen {\it et al.} \cite{ngu.19,sau.22}. As the resonance with the mode is approached, the speed of sound  increases significantly. The interference fringes were used to calculate the speed of sound which compares well with the theory of Moores and Sauls \cite{moo.93}, Fig. \ref{fig:F2}\,B and our previous work. 

After verifying the performance of the cavities in the superfluid state, and establishing a reference for $\alpha$, we performed a series of pressure sweeps in the Fermi liquid at temperatures hovering above  $T_c$. The attenuation of TZS in the Fermi liquid state is expected to decreases as $\alpha \sim T^2$, calculated in the next section and shown  in Fig.\ref{fig:F3}A. Clearly, it is best to search for  interference at the lowest temperature varying the pressure.  

\begin{figure}
\includegraphics[width=8.2cm]{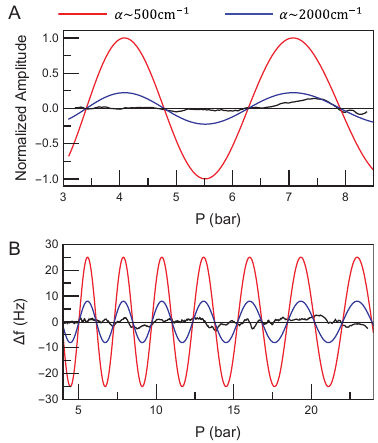}
\caption{{\bf Pressure Sweep}\\ Data in the Fermi liquid state $\sim$ 2 mK compared with the expected interference pattern obtained from numerically solving the Landau-Boltzman equation in a 5.2 $\mu$m cavity. ({\bf A}) Pressure sweep at a single fixed frequency using our CW spectrometer at fixed frequency, 103.9 MHz.
The red curve shows the expected amplitude with an attenuation of $\alpha$= 500 cm$^{-1}$. The blue curve is for $\alpha$= 2000 cm$^{-1}$. ({\bf B}) Pressure sweep of the full spectrum of the 29$^{th}$ at 143.7 MHz from which the central frequency was determined. Neither sweep indicates evidence of an interference pattern consistent with a propagating transverse mode. }
\label{fig:F4}
\end{figure}

\section{Interference Simulation}

Flowers and Richardson\cite{flo.76, ric.78} considered an oscillating semi-infinite plate with two relaxation times to account for high attenuation of TZS. Using this approach Kuorelahti \textit{et al.}\cite{kuo.16} solved the Landau-Boltzmann equation for a cavity.  Following their procedure, we  calculated the attenuation for the transverse sound   in Fig.\ref{fig:F3}A , and the imaginary part of the acoustic impedance in  Fig.\ref{fig:F3}B for a 5.2 $\mu$m cavity with frequency 103.9 MHz.

The temperature enters the zero sound regime in the Fermi liquid below 5 mK. With reduced attenuation, as shown in Fig.\ref{fig:F3}A, transverse sound is expected to propagate.  Its return to the transducer affects the acoustic impedance\cite{hal.90b} of the transducer, $Z$, (Eq. 3) producing interference fringes in the imaginary part,  $Z''$, during a pressure sweep and various temperature sweeps. To compare the numerical result with experimental data,  $Z$ is converted into a frequency shift from the relations; 
\begin{equation}\label{eq:Impedance_freq}
    Z = Z' + iZ'' = \frac{1}{4}n\pi Z_q \Delta Q^{-1} + i \left(\frac{1}{2}n\pi Z_q \frac{\Delta f_0}{f_0}\right)
\end{equation}
where $n$ is the n-th order harmonic, $Z_q$ is the impedance of quartz, $Q$ is  the Q-factor and $f_0$ is the n-th harmonic frequency.

Pressure sweeps at 21$^{st}$(103.9 MHz) and 29$^{th}$ (143.7 MHz) harmonics of the transducer are shown in Fig \ref{fig:F4}. For Fig.\ref{fig:F4}A, the RF spectrometer was tuned to a single frequency (103.9 MHz) while tracking the reflected signal from the transducer. The red and green curves show the expected amplitude and periodicity of the interference pattern for TZS with an attenuation coefficient of $\alpha$ = 500 cm$^{-1}$ and 2000 cm$^{-1}$, respectively. The amplitudes of these curves were scaled from the result in the superfluid state for the background signal in this cavity with $\alpha \sim 500$ cm$^{-1}$. It is evident that there are no interference fringes  consistent with TZS. 

For the data in Fig.\ref{fig:F4}B, a network analyzer was used to acquire the full spectrum of the 29$^{th}$ harmonic around 143.7 MHz while performing a pressure sweep. From the central frequency, both the periodicity and amplitude are directly compared with the numerical simulation. Note that the frequency shifts were derived from Eq. (\ref{eq:Impedance_freq}) Again, it is evident that there are no interference fringes. 

While it cannot be concluded that transverse sound does not propagate in the FL state, if it does exist the actual attenuation must be significantly higher than theoretically predicted within the relaxation time-approximation. At 2 mK, we conclude that the attenuation of sound is significantly greater than 2000 cm$^{-1}$ and likely to be the case at other pressures and frequencies. This is a marked extension of what has been estimated previously. \cite{roa.76, flo.76, mil.81}.

\section{ACKNOWLEDGEMENTS}
Work supported by NSF Division of Materials Research Grant No. DMR-2210112. We are grateful to J. A. Sauls and A. Vorontsov for useful discussion and thank Yoonseok Lee, John Davis, and Charles Collett for their important work on transverse sound in superfluid \he.
\bibliography{Manref}
\end{document}


\title{New Limits on Transverse Zero Sound in Fermi Liquid \he\\SUPPLEMENTARY MATERIALS}

\author{M. D. Nguyen}
\email{mannguyen2019@u.northwestern.edu,\\ w-halperin@northwestern.edu}
\author{D. Park}
\author{J. W. Scott}
\author{N. Zhelev}
\author{W. P. Halperin}
\affiliation{Department of Physics and Astronomy, Northwestern University \\Evanston, IL 60208, USA}

\date{\today}

\maketitle

\section{Transverse Sound in Superfluid \he}

\begin{figure}
\includegraphics[width=8cm]{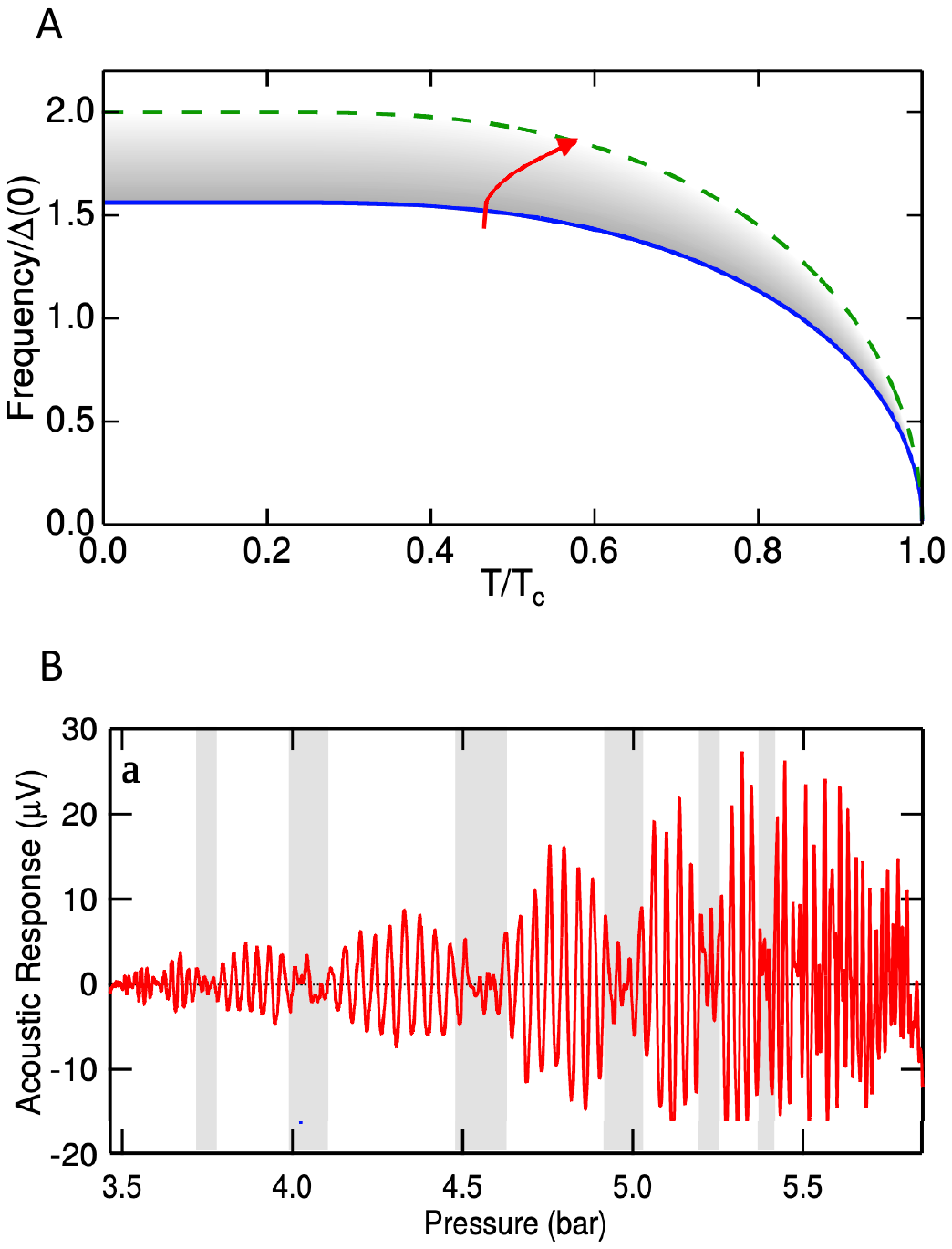}
\caption{Pressure sweep data in the \he\ superfluid state around 0.45 mK. ({\bf A}) Typical pressure sweep at a single fixed frequency  decreasing from a pressure of 6 bar. The solid blue line indicates the $J= 2-$ Higgs mode  frequency, and the green dashed line indicates pairbreaking relative to the zero temperature gap. Transverse sound propagates in the shaded region, and the red arrow indicates the sound frequency where both both pressure and temperature are changing. The curvature is caused by a combination of changes in T and P, which changes Tc. ({\bf B}) Flattened and smoothed acoustic response trace at H = 0.1 T versus pressure. Shaded regions indicate selected acoustic Faraday effect modulation minima reproduced from Collett \cite{col.15} }.
\label{fig:F1}
\end{figure}

Transverse sound was predicted by Moores and Sauls \cite{moo.93} as a collective mode in superfluid \he-B and confirmded in experiments by Yoonsek Lee {\it et al.} \cite{lee.99} using acoustic cavities. Importantly, the existence of a Faraday effect in this transverse wave propagation provides a direct manifestation of  the spontaneously broken spin-orbit symmetry which is characteristic of this superfluid phase. The experiments performed with acoustic cavities of the nominal thickness $D$\, = 30 $\mu$m provide a basis for the present results using much smaller $D$. In addition to the primary reference further experiments of key importance are the works of Davis \cite{dav.08a,dav.08b} and Collett \cite{col.15,col.13a}. In particular Collett extended the Faraday effect to magnetic fields of $H\,=\,0.1T$ for which the Faraday rotation of the polarization of the transverse sound exceeded 1,700$^\circ$ shown in Fig.\ref{fig:F1}. 

\section{Materials and Methods}

To ensure high parallelism between the two surfaces for a well-defined path length, we fabricated a 5.2 $\mu$m  cavity out of a silicon-on-insulator (SOI) substrate, Fig.F2(A). SOI wafers have 3 layers, a device layer of silicon (Si) (typically micron scale), a buried oxide (BOX) layer of SiO$_2$ (typically $\leq$ 1 $\mu$m), and finally a handle layer that is several hundred $\mu$m thick. The reflecting plate for our cavity is the Si handle layer while the spacer features are made of the BOX layer and device layer. The SOI wafer was chosen such that the device layer forms the bulk of the cavity spacing (5 $\mu$m) while the BOX layer adds 200 nm yielding a path length of 5.2 $\mu$m. The fabrication from a SOI wafer instead of simply silicon is to ensure a smooth reflecting surface, similar to the surface of the transducer. The dry-etching (Bosch) process of Si would lead to a very rough surface, not suitable for acoustic cavities. The BOX layer of the SOI wafer protects the handle layer's surface during the dry etching of the device layer. The BOX can be wet etched with hydrofluoric acid after the dry etch process, revealing the pristine handle layer. 
The transducer is an AC-cut quartz transducer that generates the transverse, thickness shear mode. with gold electrodes on the backside (non-cavity side). The fundamental frequency is 4.94 MHz and the transducer was mainly operated at  the 21st (103.9 MHz) and 29th harmonics (143.7 MHz).

However, cavities cannot be made too small since a shorter path length $D$ produces fewer interference fringes. 
The drive frequency can be increased to compensate for the reduced fringes of a small cavity but the attenuation also has frequency dependence from two sources. In the relaxation time approximation, TZS has no frequency dependence with $\alpha \sim T^2$ (unlike in the hydrodynamic limit where $\alpha \sim \omega^2/T^2$ \cite{lan.57b}).

However if the sound energy $\hbar \omega$ is close to $k_B T$, then it can excite non-thermal quasiparticles which would contribute to the attenuation \cite{lan.57b}. The crossover into what Landau calls the quantum limit of TZS corresponds to $\hbar \omega > 2 \pi k_B T$. Far into the quantum regime where $\hbar \omega >> 2 \pi k_B T$, Landau predicts a simple quadratic frequency dependence of $\alpha$. At 2 mK 100 MHz sound, $\frac{\hbar \omega}{2 \pi k_B T} \sim 0.3$,  is in the regime where some non-trivial frequency dependence is expected \cite{mat.94}. Finally if the full relaxation dynamics of the transport equation is considered, we might expect additional frequency dependence.

\section{Relaxation Times}

There is a reasonable  theoretical basis for why the attenuation may be significantly higher than expected. The Landau-Boltzmann equation (Eq.1) has a collision term that must be modeled. The simplest case is called the relaxation-time approximation where it is assumed fluctuations relax back to equilibrium exponentially with some time constant $\tau$ (i.e. $\delta I$ is proportional to the disturbance, $\frac{\Phi_{\bm{\hat{k}}}}{\tau}$). Lea \textit{et al.} assumed this to be the case yielding an approximate attenuation of 500 cm$^{-1}$ at 2 mK. You can generalize this assumption by allowing each  $l$-channel  fluctuation to relax at a separate rate. Usually this is truncated to some number of channels, such as was done by Flowers \textit{et al.} \cite{flo.78} where they considered 2 relaxation times. The relaxation time of the collective part of the oscillation is taken to be the quasiparticle relaxation time, $\tau$, and proportional to the viscous relaxation time  obtained from viscosity measurements \cite{whe.75}, $\tau_{\text{v}}$, with a parameterization 
\begin{equation}\label{eq:TwoRelax}
\tau = \xi_2 \tau_{\text{v}}.
\end{equation}
The viscous relaxation time is mostly due to the relaxation of $l = 2$ channel \cite{flo.78,ric.78}. In other words, the quadrupolar fluctuations ($l = 2$) of the Fermi surface relax with a time constant $\tau_{\text{v}} = \tau/\xi_2$. 

Estimates have been made from various experiments with $\xi_2 = 0.35$ \cite{law.73} or $\xi_2 = 0.28$ \cite{wol.76}. The smaller $\xi_2$, the greater the attenuation of sound with $\xi_2 = 0.28$ yielding about 1300 cm$^{-1}$, as seen in Fig.3. The two-relaxation time model indicates that for a given temperature, \he\ is at a lower value of $\omega \tau$ than naively expected in the one-relaxation time model. Consequently, the cavities are probing sound that is not as deep into the zero sound regime as anticipated, leading to higher attenuation. 

As seen from the multiple relaxation time model, the attenuation can vary greatly by allowing for complex relaxation dynamics. Furthermore, the value of $F_2^s$ has a significant role in determining $\alpha$. In fact, it is of similar order of magnitude as changing $\xi_2$ from 0.28 to 0.35, as shown in Fig.3 in the main text. This might account for the considerable disagreement on the value of $F_2^s$ reported from a number of experiments at low pressure\cite{ham.95}. Additional contributions  to the impedance (and correspondingly attenuation) result from consideration of vertex corrections to the quasiparticle propagator. Shahzamanian and Yavary \cite{sha.07} found that this introduces an explicit frequency dependence to the impedance. 

The conclusion we draw from our observations, previous experiments, and the various theoretical models, is that high attenuation of TZS in the Fermi liquid has prevented any direct detection of the mode from acoustic cavity interference.

\bibliography{Manref}